\documentclass{svproc}
\usepackage{url}

\usepackage{graphicx}
\usepackage{natbib}
\newcommand{\adv}{    {\it Adv. Space Res.}} 
 
\newcommand{\aap}{    {\it Astron. Astrophys.}}

\newcommand{\apj}{    {\it Astrophys. J.}}
\newcommand{\apjl}{   {\it Astrophys. J. Lett.}}

\newcommand{\jgr}{    {\it J. Geophys. Res.}}
\newcommand{\jgra}{    {\it J. Geophys. Res. Atm.}}
\newcommand{\lrsp}{    {\it Liv. Rev. in Solar Physics}}

\newcommand{\planss}{  {\it Planetary Space Science}}

\newcommand{\solphys}{{\it Solar Phys.}}
 
\newcommand{\ssr}{    {\it Space Sci. Rev.}} 

\begin{document}
\mainmatter              
\title{CME Observations – from Sun to Impact on Geospace}
\titlerunning{CMEs and their Influence on Heliosphere and Earth}  
%
\author{Manuela Temmer\inst{1}}
\authorrunning{M. Temmer} 
%
 
%
\institute{Institute of Physics, University of Graz, Austria\\
\email{manuela.temmer@uni-graz.at},\\ WWW home page:
\texttt{https://swe.uni-graz.at}
}

\maketitle              

\begin{abstract}
Our Sun is an active star expelling dynamic phenomena known as coronal mass ejections (CMEs). The magnetic field configuration on the Sun and related solar wind structures affect the propagation behavior of CMEs, dominate its transit time and embedded magnetic field properties when impacting Earth. Since the conditions on the Sun constantly change, the impact of CMEs on the different regimes of geospace is quite variable and may differ significantly from event to event. This short review summarizes the different manifestations of CMEs on the Sun, their appearance in interplanetary space, and how CMEs trigger a cascade of reactions as they interact with Earth. 
\keywords{Coronal mass ejections, Solar Wind, Interplanetary Space, Earth, Space Weather research}
\end{abstract}

\section{Introduction}

\begin{figure*}
   \centering
   \includegraphics[width=\textwidth]{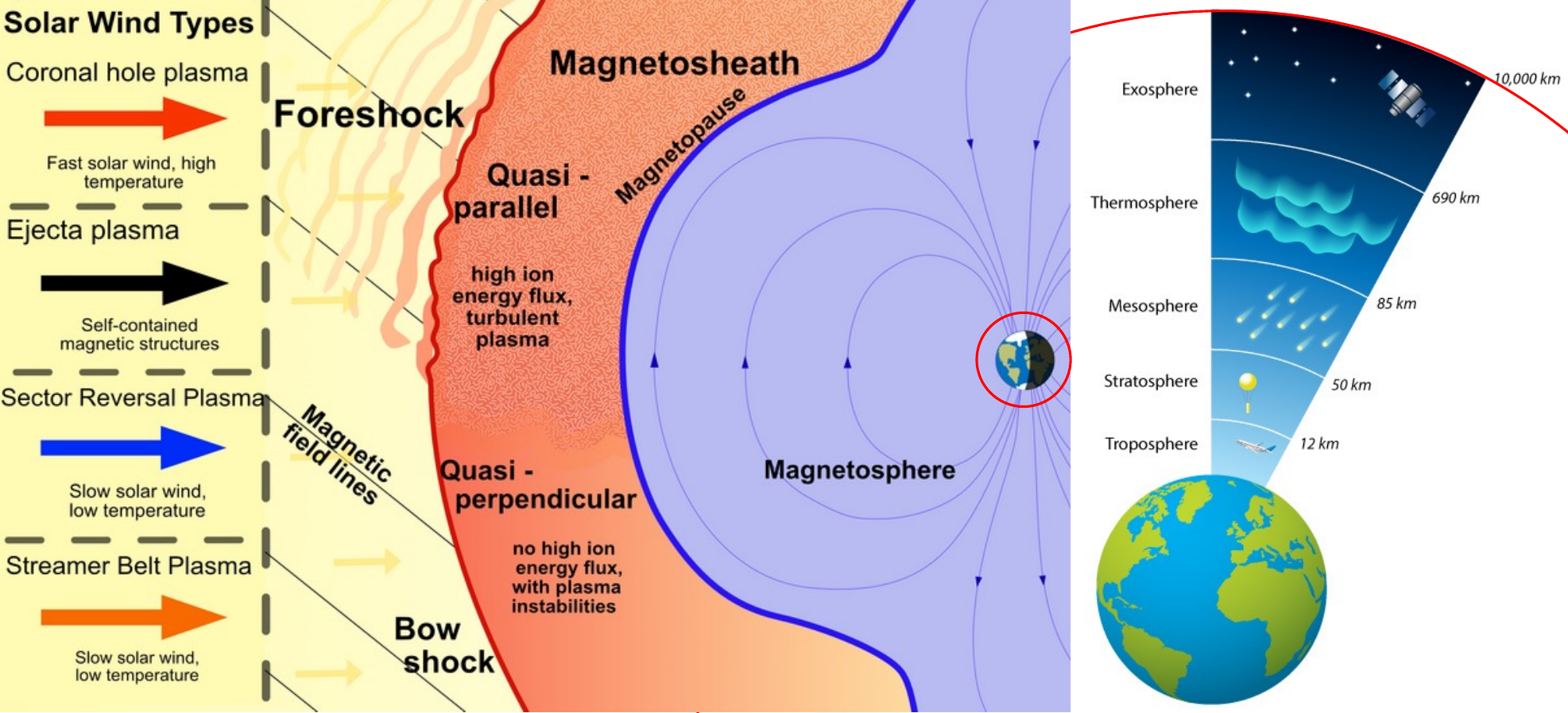}
      \caption{Left: Graphic representation of the boundary region between the solar wind and the Earth's magnetic field  \citep[adapted from][]{koller2024}. Right: The atmospheric layers of Earth - a zoom in version of the red circle marked in the left panel (source: European Centre for Medium-Range Weather Forecasts).
}
         \label{impact}
   \end{figure*}

Coronal mass ejections (CMEs) are the most energetic events from the Sun, appearing as rapidly moving and expanding magnetic plasma structures. Frequently CMEs are associated with flares \citep{Yashiro2006}, which actively drive CMEs, as well as solar energetic particles (SEPs) which are accelerated by the huge energy release in the wake of the magnetic reconnection process and by the shock that is driven by the CME \citep{Donald13}. CMEs, flares, and SEPs are recognized as key factors in investigating Space Weather \citep[e.g.,][]{LuhmannEtAl2020,Temmer2021b,Gopalswamy2022}. Out of that CMEs are known to cause the most severe Space Weather effects, such as geomagnetic storms that can induce electric currents in power lines, potentially leading to widespread electrical grid failures and damage to infrastructure \citep{Pulkkinen17}. Intense research for better understanding the physical processes in the complex and interdisciplinary area of Space Weather is therefore essential for safeguarding technology, infrastructure, and human activities both in space and on Earth (see Figure~\ref{impact}). 

There are many open questions, e.g., about the interaction processes between CMEs and the ambient solar wind flow. Especially interaction of CMEs with other large-scale solar wind structures lying in the path of the CME, produces complex interplanetary magnetic field structures significantly enhancing Space Weather effects \citep[e.g.,][]{Dumbovic15}. Most of all, we face the fact that the ``plain'' background solar wind and its embedded structures, such as stream interaction regions and high-speed solar wind streams, are not well represented by state-of-the-art models \cite[see e.g., COSPAR Space Weather Roadmap update by][]{Temmer23_arxiv}. This lowers the reliability of Space Weather forecasts, especially during high solar activity phases when interactions might happen frequently \citep{Lugaz2017}.

\section{Towards more reliable background solar wind models}

It is acknowledged that the open magnetic flux (OMF) ``shapes'' interplanetary space, magnetically connects the Sun to the planets, and dominates the motion of SEPs \citep[e.g.,][]{Owens2013b}. Coronal holes (CHs) are assumed to mainly represent the OMF \citep[e.g.,][]{Hofmeister2018}. CHs are also the sources of fast solar wind streams which generate, when interacting with the ambient slow solar wind, so-called stream interaction regions (SIRs). A CME encountering a SIR and related high-speed stream may experience acceleration, changes in the magnetic field configuration and direction of motion, resulting in altered transit times and geomagnetic effects at Earth.

Investigating the dynamic interplay between open (fast wind) and closed (slow wind) magnetic field on the Sun is key in deriving answers on how interplanetary space is structured. However, there is a strong discrepancy found when calculating the in-situ OMF and the OMF coming from the solar surface with an underestimation by a factor of two \citep[e.g.,][]{Linker2017}. If we assume the accuracy of the in-situ OMF calculation, any unaccounted-for OMF must be sought at the Sun. There might be outflows from active regions that could contribute to the OMF \citep[e.g.,][]{vanDriel2012}. Recent studies derived that the OMF evolution correlates with CH open flux, but not with CH area evolution \citep{Heinemann2024}. Hence, the uncertainty in detecting boundaries of CHs might not be the crucial point in the missing OMF and, hence, might not be the main parameter in improving the performance of background solar wind models \citep{Linker2021}. Rather, the currently missing polar magnetic field information might play a key role. ESA's mission Solar Orbiter \citep[SolO;][]{Mueller2017} will move out of the ecliptic plane starting in 2026. It will deliver high-resolution magnetic field information from the Sun's polar region, which might enable us to get one step closer towards solving the OMF problem and more reliable background solar wind models.

\section{Solar signatures related to CMEs}

The regions on the Sun from which CMEs originate give limitations in the driving force of a CME \citep{Bein2011}. Active regions with strong magnetic fields and complex sunspot configurations are often associated with greater solar eruptive potential \citep[e.g.,][]{Sammis00}. The common ground for the occurrence of CMEs and flares is magnetic reconnection and magnetohydrodynamic instability, however, recent studies highlight that the trigger mechanisms for CMEs can vary \citep[see review by][]{green2018}. 

\subsection{Signatures of eruptive events}

Besides the high association rate between flares and CMEs \citep{Yashiro2006}, there are also well-known active regions producing confined flares of high energy release \cite[e.g.,][]{thalmann2015}. While confined flares can still produce intense bursts of electromagnetic radiation across various wavelengths, they do not lead to CMEs. It is therefore important to be able to distinguish confined from eruptive flares. 

For that radio signatures are favorable. Especially the detection of type II and type III radio bursts are important indicators of solar eruptive events. Type II radio bursts are related to fast CMEs, and are indicative of the passage of shock waves through the solar corona and interplanetary space \citep[see e.g.,][]{Gopal2008}. 
Type III bursts are characterized by rapidly drifting frequency signatures as the electron beams propagate away from the Sun along magnetic field lines and are indicative of the opening of the magnetic field. Hence, the observation of type III followed by type II bursts can be used to infer the eruptive and energetic nature of a flare. 
E.g., for continuously monitoring the Sun's radio emissions, the e-CALLISTO \citep[Enhanced Coordinated Low-cost Low-frequency Instrument for Spectroscopy and Transportable Observatory;][]{Benz2009} network consists of multiple stations distributed worldwide, each equipped with a radio spectrometer ranging typically from a few tens of megahertz to a few gigahertz.


Before entering the field-of-view of the coronagraph, typically solar surface signatures collectively provide evidence of the occurrence and characteristics of eruptive flare events and CME characteristics. It is therefore important to monitor the Sun especially in EUV or X-rays from which, together with magnetic field information, we can assess the energy storage and release in the corona. The close connection between the energy release of flares and CMEs in various wavelengths gives us an estimate of the driving forces of a CME \citep{ZhangDere2001,Temmer2008}. EUV dimming regions and their dynamics tell us about the actual mass release, the CME speed as well as the CME's direction of motion \citep{Dissaueretal2018,Chikunova23}. Propagating surface waves show us the expansion behavior of the CME \citep[e.g.,][]{Patsourakosetal2020}. Post-eruptive arcades can be used as proxy for the magnetic structure of the flux rope and the amount of reconnected flux, frequently used in models \citep[e.g.,][]{Scolini2020a}. The coronal hole influence parameter (CHIP) determines the influence of nearby CHs on the CME trajectory \citep{Gopal2009}.



\section{CME structures and their differences}

Focusing on the predictability of Space Weather, we first need to consider the different structures of a CME. In rough terms, we may differentiate between the shock-sheath and the magnetic ejecta of a CME \citep[e.g.,][]{Vourlidas2013}. Based on statistical results, the difference in the arrival time of the shock-sheath and magnetic ejecta is on average about 10 hours \citep{Russell2002}. Geomagnetic effects associated with the shock-sheath include sudden increases in solar wind speed, pressure, and magnetic field intensity, leading to enhanced geomagnetic activity, which are often short-lived but can be intense. 
The CME magnetic structure plays a more crucial role in enhanced geomagnetic effects as a southward directed magnetic field component (negative B$_z$, which is opposite to Earth's magnetic field), can efficiently couple with Earth's magnetosphere and trigger magnetic reconnection \citep[e.g. review by][]{Tsurutani2023}. In that respect we may ask ourselves if we are consistent in our understanding of different CME structures and how well we are able to identify them in remote-sensing image data and to track them from Sun to Earth \citep[see e.g.,][and related ISSI team]{Verbeke2023}.

\begin{figure*}
   \centering
   \includegraphics[width=\textwidth]{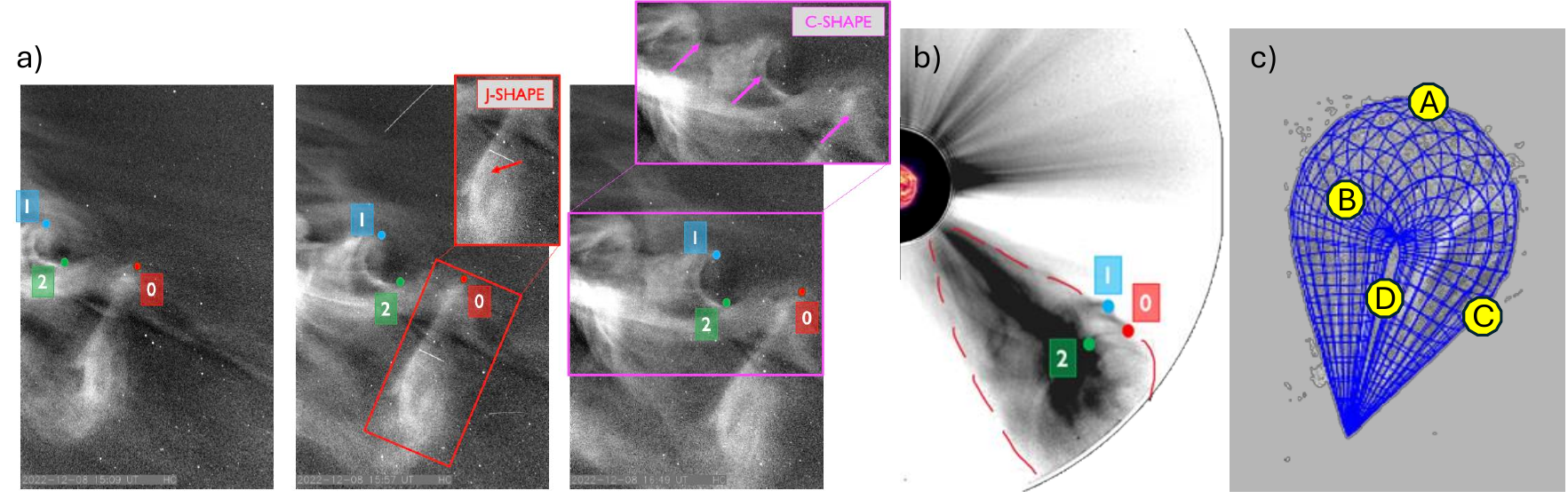}
      \caption{a): WISPR-I white-light data from 2022-12-08 showing in detail some identified small-scale internal CME structures (0, 1, 2). b): STEREO-A COR2 coronagraph image, showing the tracked features in the global view. Adapted from \cite{cappello24}. c): Cartoon of a GCS reconstructed CME with regions of interest highlighted: front (A), internal structure (B), flank (C), and current sheet (D). 
}
         \label{cme_detail}
   \end{figure*}

White-light remote sensing image data give us a global view on the CME, whereas in-situ measurements reveal local variations. The closeness of Parker Solar Probe \cite[PSP;][]{Fox2016} and SolO spacecraft to the Sun, together with well established and newly developed triangulation methods, enable us to investigate for the first time small-scale structures at different regions of the CME (see Figure~\ref{cme_detail}). Results from \cite{cappello24} revealed that blob-like density structures distributed in the rear-part of the CME,  might reveal the interaction process with the ambient solar wind. This helps us in better understanding the propagation behavior of CMEs in interplanetary space as well as to connect globally and locally derived measurements. 

\section{Imprint of CMEs as they propagate through the heliosphere}

\begin{figure*}
   \centering
   \includegraphics[width=0.8\textwidth]{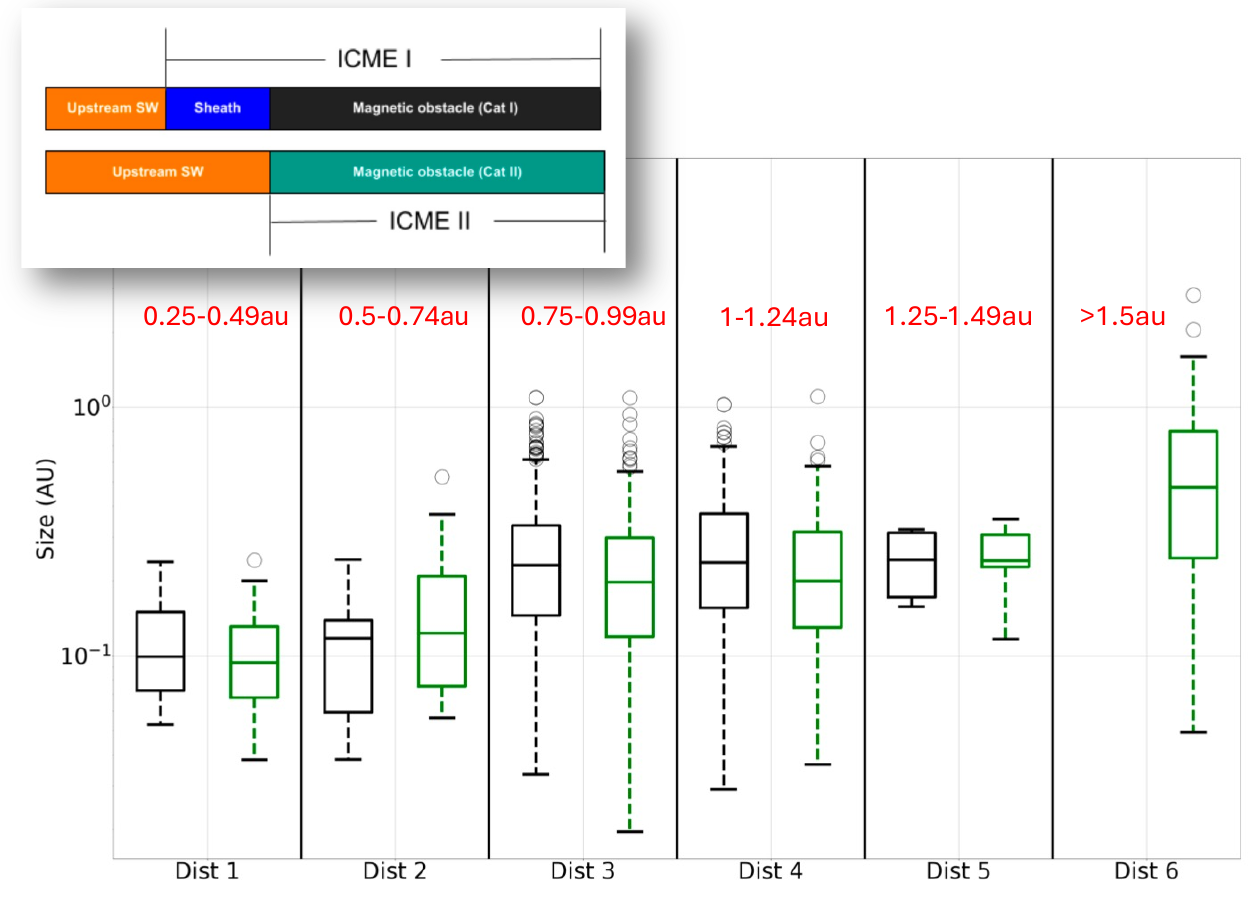}
      \caption{Size of CME magnetic obstacles driving a clear sheath (black) and those w/o sheath (green). The size (mean speed over structure $\times$ duration) is calculated for measurements derived at different distances from the Sun covering the inner heliosphere to 1 AU. Adapted from Larrodera \& Temmer, 2024.
}
         \label{carlos}
   \end{figure*}

As the evolving CME compresses the solar wind ahead of it, it forms turbulent high-pressure regions known as sheath, and if fast enough, these steepen up into shocks. The sheath is a high plasma-beta region, hence, it holds implications for the draping of the interplanetary magnetic field and accumulation of plasma \citep{Siscoe2008}. How the sheath region might affect the expansion behavior of the CME's magnetic structure and how it is related to interplanetary magnetic field draping is another open issue. A recent comprehensive study on CMEs with and without clear sheath regions over the inner and outer heliosphere is presented by \cite{Larrodera2024}. The dataset covers more than 2000 CMEs over the time range 1975--2022. As can be seen in Figure~\ref{carlos}, it is found that both types (CMEs with and without sheath) increase in size from the inner heliosphere to 1 AU by ca. 47\%, with the strongest size increase around 0.75\,AU. There were no CMEs found with sheath structures ahead beyond 1.5\,AU. CMEs without sheath seemed to almost double their size beyond 1.5\,AU compared to 1\,AU, however, the data sample is rather low for the outer heliosphere. A review on CME sheaths is given by \cite{Kilpua2017a} and there are interesting aspects found in CME sheath regions as shown in other studies \citep[e.g.,][]{Salman2021}. 



During high solar activity we frequently observe multiple CMEs (see e.g., episodes of enhanced geomagnetic effects due to multiple CMEs in March, April, and November 2023, or most recent in May 2024). In the late phase of solar activity, coronal hole structures at low latitudes are frequently observed, and SIRs dominate in interplanetary space. These multiple eruptions can lead to the formation of very complex magnetic structures in the heliosphere significantly complicating the prediction of their impact on Earth \citep{Gopalswamy2001, BurlagaEtAl2002, Harrison2012} and causing to the most intense geomagnetic storms \citep{Farrugiaetal2006,xie04,Dumbovic15}. Besides the mutual interaction between the various structures, each CME (also SIR) generates perturbations in the smooth outflow of the slow solar wind and changes the magnetic field configuration. With that interplanetary space and near-Earth conditions experience a preconditioning effect that might last for several days \citep{Temmer2017,Janvieretal2019}. The occurrence of strong eruptions can also lead to transient open field conditions making subsequent CME propagation super-fast \citep{Liu2014}.

\section{Geospace Impact and Interdisciplinary Research}

Due their complex nature, the detailed impact of CMEs on geospace is highly variable. When CMEs collide with Earth, they trigger a cascade of reactions which are started at the magnetosphere. These reactions include the opening of the magnetic field, compression, and the occurrence of substorms in the magnetotail \citep[see e.g.,][]{Russell73,Gosling91}. The strongly varying solar and magnetospheric energy input leads to significant fluctuations in the Earth's ionosphere \cite[see e.g.,][]{Buonsanto99,Tsurutani2004}. The thermosphere gets heated and expands due to the absorption of EUV and X-ray radiation and, as the CME reconnects with the magnetosphere, due to injection of additional particles into that atmospheric region. This leads to an increase in the neutral density in larger heights of the atmosphere, leading to stronger drag for low Earth orbiting (LEO) satellites \cite[see e.g.,][]{Knipp2004,Krauss2015,Bruinsma2023}. 

Recent studies found that density enhancements in the magnetosheath, so-called magnetosheath jets (MJ), represent a significant coupling effect between the solar wind and the Earth's magnetosphere \citep[see review by][]{Plaschke2018}. A significant variation in the number of MJs is derived with CMEs/SIRs actually lowering/enhancing the MJ production rate \citep{Koller2022}. More interdisciplinary research is needed to better understand the coupling processes.

\section{Conclusion}

CMEs are pivotal in shaping Space Weather, influencing both the heliosphere and geospace. The driving forces of CMEs are related to the conditions on the Sun that lead to CME formation. The factors influencing their propagation through the heliosphere are likewise related to slow and fast solar wind regions in terms of CHs. Knowledge on the surface structures and how these further evolve into interplanetary space is therefore crucial for accurate forecasting. Small scale structures embedded in the CME or signatures of solar wind interaction, need yet to be studied to find out more about their effects on Space Weather. 

Forecasting the impact of CMEs, however, remains challenging due to their complex and variable nature, the not-well understood interaction processes with the solar wind, as well as lack in accurate knowledge of the background solar wind itself. Advanced coronal and heliospheric models incorporating real-time solar wind data and detailed magnetic field measurements are essential for improving predictions. The  unprecedented data gathered by PSP and SolO, currently provide and will provide detailed information of the near-Sun environment, enhancing our understanding of CME initiation and interaction processes with the solar wind. 

At the impact of the various solar wind structures on Earth, the cascade of processes in the magnetosphere and lower atmospheric layers is not fully understood. More interdisciplinary research is therefore needed to identify universal processes and to tackle problems from different aspects.

%


%
%
\bibliographystyle{styles/bibtex/spbasic.bst}

\end{document}